\documentclass[12pt]{article}
\usepackage{graphics}
\usepackage{graphicx}
\usepackage{amsmath}
\usepackage{amssymb}
\usepackage{bm}

\begin{document}

\title{Controlled Anisotropic Deformation of Ag Nanoparticles by Si Ion Irradiation}
 \author{A. Oliver, J. A. Reyes-Esqueda, J. C. Cheang-Wong,\\C. E. Rom\'an-Vel\'azquez,A. Crespo-Sosa, L. Rodr\'iguez-Fern\'andez, J. A. Seman,\\and Cecilia Noguez\thanks{E-mail: cecilia@fisica.unam.mx\hskip1cm Phone: +52 55 56225106\hskip1cm Fax: +52 55 56161535}\\
\it Instituto de F\'{\i}sica, Universidad Nacional Aut\'onoma de M\'exico,
 \\ \it Apartado Postal 20-364, D.F. 01000,  M\'exico, and\\ \it
 REGINA: The  Nanoscience Research Network at UNAM.}

\maketitle

\begin{abstract}
The shape and alignment of silver nanoparticles embedded in a glass matrix is controlled using silicon ion irradiation.  Symmetric silver nanoparticles are transformed into anisotropic particles whose larger axis is along the ion beam. Upon irradiation,  the surface plasmon resonance of symmetric particles splits into two resonances whose separation depends on the fluence of the ion irradiation. Simulations of the optical absorbance show that the anisotropy is caused by the deformation and alignment of the nanoparticles, and that both properties are controlled with the irradiation fluence. 
\end{abstract}

 

Nanosized metal particles exhibit optical surface plasmons that have revelead new aspects for applications in photonics devices \cite{barnes,maier}. Surface plasmons are tailored by altering the surface of nanoparticles (NPs) in some way \cite{3,4}. The potential applications of NPs depend on the ability to control their size, shape, and environment. Thus, new synthesis methods have been developed, like the ion implantation technique, which is very versatile due to the possibility of fabricating both, metal and semiconductor NPs embedded in a variety of host matrices \cite{nalwa,8}. This method has advantages over other techniques, because of the high control on the distribution and concentration of impurities, allowing the growth of NPs in a well-defined space region of the host matrix, such that, buried channel waveguides are directly formed  \cite{nalwa}. Besides, high concentrations of metal NPs can be reached, yielding values for the third order nonlinear optical susceptibility much larger than those found in other metal-doped solids \cite{10}. On the other hand, to obtain a second order nonlinear optical response, a macroscopic noncentrosymmetry must be induced in some way, since nanocomposites are isotropic and centrosymmetric. Up to now, only very small second harmonic generation signals have been measured \cite{mochan}. 

Anisotropic deformation of silica using MeV ion irradiation has been observed  \cite{18}. Also, ion irradiation in the keVÐ- MeV energy range for micrometer-sized colloidal silica particles was employed to change their shape from spheres into oblate spheroids \cite{19,192}. In the same way, gold NPs surrounded by a silica shell irradiated with 30 MeV Se ions have been deformed \cite{20}. In this case, the spherical shell was deformed into an oblate spheroid, such that, its major axis was in the direction perpendicular to the ion beam, while the Au core was deformed into a nanorod in the parallel direction. However, this effect has not been observed in pure metallic NPs, although some attempts have been done with silver: Ag NPs embedded in a planar sodalime glass film were irradiated with 30 MeV Si ions, producing chains of NPs aligned along the ion beam direction, but no shape deformation was observed \cite{13,14}. Despite of the large effort done up to date \cite{18,19,192,20,13,14}, rigorous procedures to control the deformation of embedded metal NPs using ion beam irradiation have not been established.

 In this letter, we show that the ion irradiation not only induces anisotropic deformation of silver NPs, but also that the deformation rate is controlled with the fluence of the ion beam. Theoretical simulations of the optical absorption data show that the polarization dependence of the surface plasmons is in agreement with models for prolate silver NPs. High-resolution transmission electron microscopy (HRTEM) also shows that the NPs do not form chains in any direction. Therefore, we conclude that the anisotropic optical response of the nanocomposite is attributed only to the deformation and alignment of the silver NPs, which are controlled with the ion beam fluence.

\begin{figure}[htbp] 
   	\centering
	\includegraphics[]{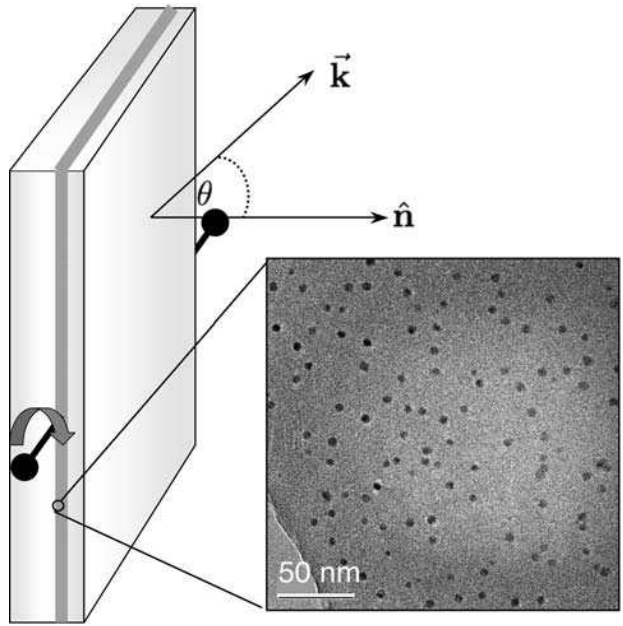} \vskip-.1cm
   	\caption{Schematic model of the nanocomposite and a HRTEM micrograph. 
}
   \label{fig1}
\end{figure}

Using high-purity silica glass plates ($20\times20\times1$mm$^3$), NSG ED-C (Nippon Silica Glass) as  host matrices,  Ag nanoparticles are synthesized  by implanting  2~MeV Ag$^{2+}$ ions at room temperature. The system is then thermally annealed at $600^\circ$C in a 50\%N$_2+$ 50\% H$_2$ reducing atmosphere. The Ag-ion fluence and projected range are of $5\times10^{16}$ Ag/cm$^2$ and $0.9$ $\mu$m, respectively, as measured by Rutherford backscattering spectrometry (RBS). Ion implantation and RBS analysis were performed at the 3 MV Tandem accelerator (NEC 9SDH-2 Pelletron). Optical absorption spectra show the formation of silver NPs \cite{8}, with a single narrow surface plasmon located at 400~nm, which is characteristic of spherical-like shaped NPs \cite{4}. Afterwards, the silica plate is cut in pieces and each piece is irradiated at room temperature with 8 MeV Si ions. According to our previous results concerning the ion beam-induced deformation of silica particles \cite{22}, 8 MeV Si ions were chosen since their electronic stopping power for SiO$_2$ is 200 times higher than the nuclear one. On the other hand, the ion projected range for this energy is 4.3 $\mu$m in SiO$_2$, i.e., far beyond the location of the Ag NPs.  
The Si irradiation was performed under an angle off normal of  $\theta = - 41^\circ$. Each sample is irradiated at different Si fluences in the range of $0.1$ -- $2.0 \times10^{16}$ Si/cm$^2$  in order to induce deformation in the Ag NPs. 
In the left-hand side of Fig.~\ref{fig1}, a schematic model of the buried region into the glass sample containing the silver NPs is shown. The incident ion beam makes an angle $\theta= - 41^\circ$ with the normal $\hat{\mathbf n}$. In the right-hand side a HRTEM micrograph of the in-normal view of the sample shows that the silver NPs are randomly located, i.e., they do not form chains or any other kind of arrangement. The HRTEM study was carried out in a JEOL 2000F at 200 kV, by uncapping the sample  in the direction of the surface normal until few hundreds of nanometers, employing an ion beam milling. The size distribution was obtained from a digitalized amplified micrograph by measuring the diameter of each NP. We perform a statistic over 290 NPs, and we find a diameter distribution  centered at 5.9~nm  with a standard deviation of 1.1~nm. 

\begin{figure}[hbpt] 
   	\centering
	\includegraphics[]{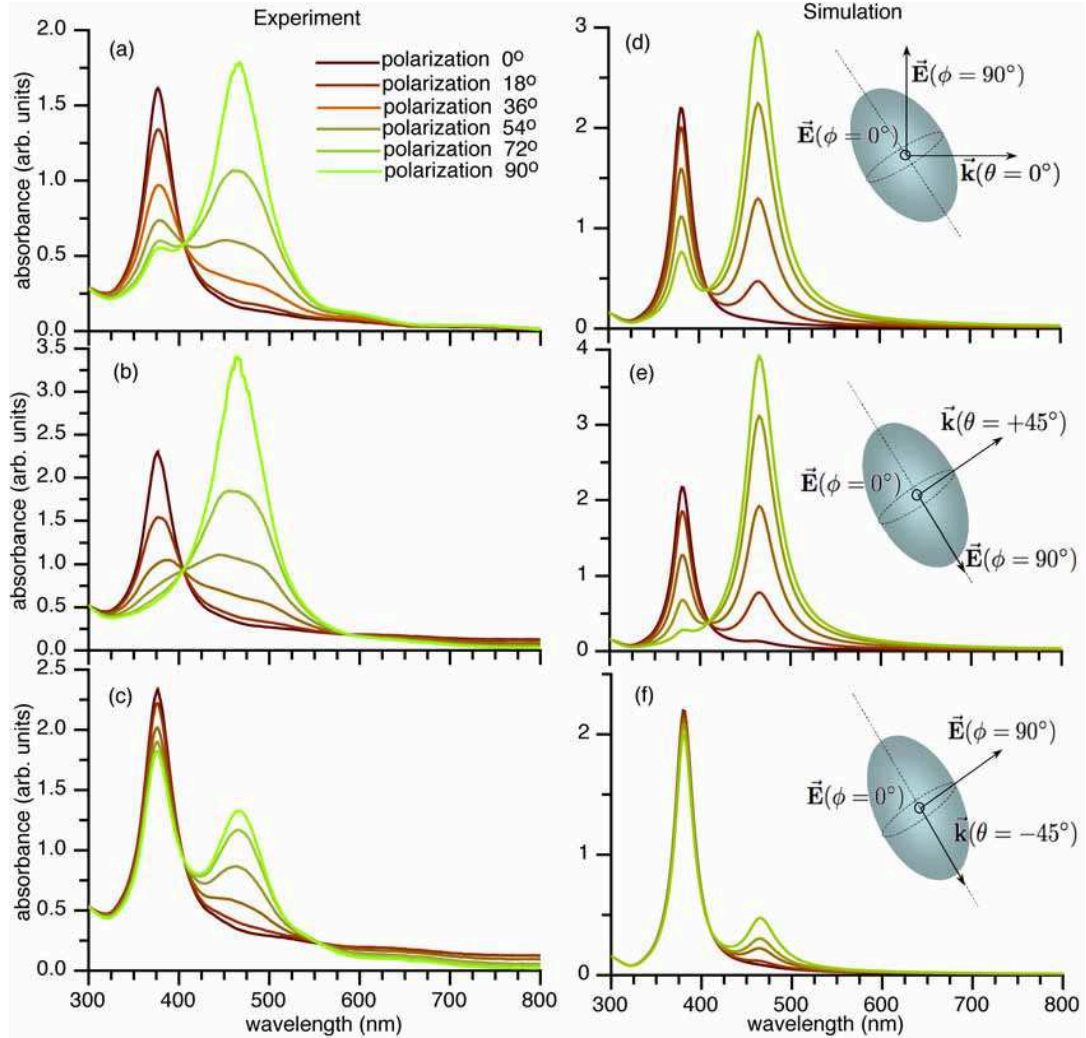}\vskip-.1cm
   	\caption{Measured (left) and simulated (right) optical absorbance  for linear-polarized light from $0^\circ$ to $90^\circ$, and (a), (d) $\theta=0^\circ$, (b), (e) $\theta= + 45^\circ$, and (c), (f) $\theta= - 45^\circ$. The polarization is indicated in the first plot, and is the same for all plots. In the simulated spectra the orientation of the NP's major axis with respect to the incident electromagnetic field is depicted.}
   \label{fig2}
\end{figure}

The optical absorbance of the nanocomposites has been measured by controlling the angle $\theta$ of the wavevector $\vec{\mathbf{k}}$ of the incident electromagnetic field with respect to $\vec{\mathbf{n}}$, as shown in Fig.\ref{fig1}. The polarization of the incident electric field was varied at different angles $\phi$ with respect to the horizontal axis passing through the middle of the plane of the sample and perpendicular to $\vec{\mathbf{k}}$. An Ocean Optics Dual Channel S2000 UV-visible spectrophotometer was used to collect the electronic absorption spectra.  In Fig.~\ref{fig2} we present the measured (left) and simulated (right) absorbance for a nanocomposite after Si irradiation with a fluence of $0.5 \times10^{16}$ Si/cm$^2$. In Fig.~\ref{fig2} (a) the experimental absorption spectra for  $\theta = 0^\circ$ and different angles of polarization are shown.  When $\phi = 0$ in Fig.~\ref{fig2} (a), a surface plasmon resonance at about 375~nm is observed. As the angle of polarization increases, the intensity of this first resonance decreases and a new peak appears at about 470~nm and becomes more intense,  dominating the spectra at $\phi=90^\circ$. Notice that the first resonance is small but not null when $\phi=90^\circ$. 

Figs.~\ref{fig2}(b), and (c) show the absorbance of the same composite, but with $\vec{\mathbf{k}}$ at $\theta= + 45^\circ$ and $ - 45^\circ$, respectively. When $\theta=  + 45^\circ$ in Fig.~\ref{fig2} (b), the spectra show the same behavior as described above for $\theta=0^\circ$, except that when the polarization angle is $\phi =90^\circ$, the first resonance at 375~nm completely disappears. Conversely in Fig.~\ref{fig2} (c), when the incidence angle is $\theta = - 45^\circ$ the resonance at 375~nm is present for any $\phi$, and its intensity shows small variations. In this case, the resonance at 470~nm is always less intense, and for $\phi = 0^\circ$ is null. From the experimental spectra, we observe that the resonance at 470~nm disappears always for one polarization ($\phi=0^\circ$), for any angle $\theta$ of the incident electromagnetic field. All these results mean that the optical response of the system is anisotropic and has only one axis of symmetry. Furthermore, this axis should be along the direction related to the resonance at 470 nm. It is concluded that this optical behavior is qualitatively reproduced and explained by simply considering the NPs as prolate spheroids.

To corroborate the last statement, the optical absorbance spectra were simulated by employing prolate spheroids whose symmetry axis is along the major axis, as depicted in Fig.~\ref{fig2} (d), (e), and (f).  The main interest are silver NPs whose sizes are smaller than 10~nm, and in this case, radiation effects, such as scattering and radiation damping, are negligible  \cite{4}. Then, the optical absorbance was simulated by calculating the induced polarization within the quasi-static approximation, for spheroids with different aspect ratio among its major and minor axes \cite{3}, and employing a dielectric function measured for bulk silver \cite{johnson}. The macroscopic dielectric function has contributions from interband and intraband electron transitions, such that, below 320 nm, the light absorption is mainly due to the interband electronic transitions of silver. This feature is quite independent of the shape and size of the NPs, as it is actually corroborated in all the experimental spectra.  In contrast, a Drude like model of free electrons  describes the intraband electron transitions. Nevertheless, we have to consider that conduction electrons suffer an additional damping effect due to surface dispersion or finite size effects \cite{4}. 
Besides, the NPs concentration on the matrix is small, so the interactions between NPs are negligible. As a result of all these considerations, it is valid to model the absorbance of the nanocomposite as the optical response of one embedded Ag NP weighted  with the particleÕs concentration.


It was found that prolate spheroids with an aspect ratio of 1.615 and a major axis of 8 nm reproduce very well the main features of the experimental optical spectra. This aspect ratio explains why it is difficult to observe the deformation in the micrographs at first glance, since they are taken at an angle of about $ - 41^\circ$ from the major axis of the NPs. In this case, the projected size of about 1.16 coincides with the small differences between the larger and smaller axes observed with HRTEM.  In any case, it is difficult to conclude that NPs are deformed from the micrographs only. In Fig.~\ref{fig2} (d), (e) and (f) it is shown the simulated absorbance spectra for the incident electromagnetic field at  $\theta= 0^\circ$, $+45^\circ$ and $-45^\circ$, and the same angles of polarization as in the experiments. When $\theta=0^\circ$ in Fig.~\ref{fig2} (d), it is observed that  $\vec{\mathbf{k}}$ makes an angle of about $50^\circ$ with the major axis, and for $\phi=0^\circ$ the electric field is along the minor axis exciting only the surface plasmon at 375 nm. Conversely, when the angle of polarization is  $\phi=90^\circ$,  both resonances are excited, but the one at 375 nm is weaker than the resonance at 470 nm.  Similarly, when  $\theta =  + 45^\circ$ in Fig.~\ref{fig2} (e), the wavevector $\vec{\mathbf{k}}$ is almost aligned to one of the  minor axes, in consequence, the electric field is polarized along the other minor axis at  $\phi=0^\circ$, and along the major axis at $\phi=90^\circ$. Finally, when  $\theta = -45^\circ$ in Fig.~\ref{fig2} (f), the wavevector $\vec{\mathbf{k}}$ is along  the major axis, in such a way that the electric field mostly excites the resonance at 375 nm for any polarization. Since $\vec{\mathbf{k}}$ and the major axis are not completely aligned, because the implantation was done at an angle $ - 41^\circ$ off-normal, small contributions from the second resonance at 470 nm are observed.  The simulated spectra are in very good agreement with the experimental measurements. Therefore, it is concluded that silver NPs suffer a deformation due to MeV Si ion irradiation that can be explained in terms of aligned prolate spheroids.

The optically observed NPs deformation is also in agreement with a classical viscoelastic model, where it is assumed that, due to the high electronic stopping power of the impinging Si ions, a cylindrically shaped region around the ion track is subject to transient heating producing a thermal spike \cite{25}.  When a high-energy ion penetrates a material, it loses energy by ionization events and atomic collisions. It is well established that the deformation is mainly driven by electronic excitations rather than atomic displacements induced by the ion beam \cite{192}.
In such case,  the radiation induces a plastic flow that causes macroscopic stress relaxation in the material.  This plastic flow is anisotropic \cite{25} and, for Au NPs embedded in glass, it has been found  that the deformation takes place along the ion beam \cite{lamarre}, while for amorphous materials the deformation is along the perpendicular direction \cite{192}. Recently, a detailed analysis of this stress relaxation process has been reported \cite{26}. The deformation of the NPs into a prolate-like shape is also in agreement with theoretical and experimental results of the morphology dependence of Ag NPs with size~\cite{baletto05}. It has been found that Ag icosahedral nanoparticles change their morphology to decahedral or octahedral particles as their size increases, acquiring anisotropy  shapes as prolate-like particles. In our case, the volume of the sphere-like Ag NPs increases about twice upon the Si irradiation. It might be that the plastic flow and thermal spikes increases the mobility of the Ag NPs, favoring the aggregation of two neighbor NPs along the ion track during the implantation. This mechanism favors the anisotropy growth of the NPs.

\begin{figure}[htbp] 
   	\centering
	\includegraphics[]{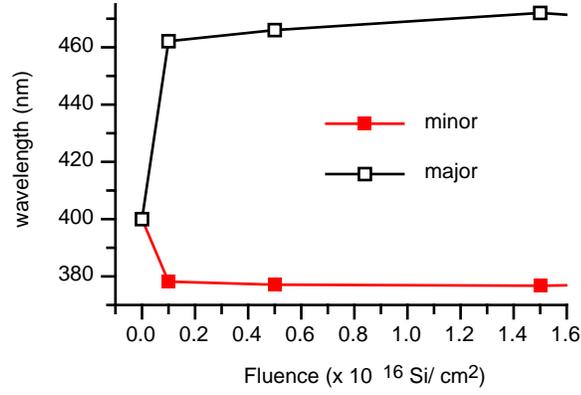} \vskip-.1cm
   	\caption{Plasmon resonance position as a function of the Si-ion fluence. The measurement was made with $\theta= +45^\circ$, and  light polarized at $\phi= 0^\circ$ and $90^\circ$.  }
   \label{fig3}
\end{figure}

When the fluence of the Si-ion irradiation varies systematically from  $0.1$ to $2.0 \times10^{16}$ Si/cm$^2$, the behavior of the absorbance for all  nanocomposites, as a function of the angles of incidence  $\theta$ and polarization $\phi$, is the same as the one described above. However, as it is shown in  Fig.\ref{fig3}, as the Si-ion irradiation fluence increases, the resonance below 400 nm shifts to lower wavelengths, while the resonance above 400 nm shifts to larger ones. This separation between resonances is also explained in terms of aligned prolate spheroids: as the aspect ratio between major and minor axes increases, the separation among resonances also does. Therefore, as the irradiation fluence increases, it is concluded that the NPsÕ anisotropic deformation increases as well. 

\begin{figure}[htbp] 
   	\centering
	\includegraphics[]{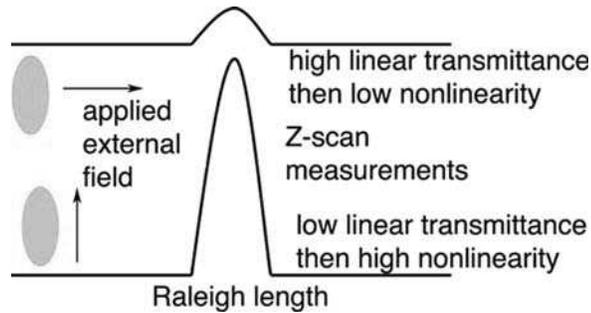} \vskip-.1cm
   	\caption{Schematic model of the third order nonlinear optical response of the deformed NPs.}
   \label{fig4}
\end{figure}

We also have experimental evidence of a rather unusual third order nonlinear optical effect on this kind of samples. In  Fig.\ref{fig4} is shown a scheme of the measured nonlinear response, using  nano and femtosecond pulsed beams at a wavelength of 532~nm, which is near the major axis resonance, where an open aperture z-scan was performed. Outside of the Raleigh region, the linear transmittance, which depends on the polarization of the incident beam, is dramatically lower for the electric field parallel to the major axis of the NPs than in the opposite case. The latter also proves the NPs deformation. Inside of the Raleigh region, a dramatic saturation of the linear absorption (nonlinear saturable absorption) at a very low energy of the light-beam is found when the electric field is parallel to the major axis. This is also observed for the opposite polarization but at rather large energies. A more detailed analysis of the nonlinear optical properties of these systems will be discussed elsewhere. 

In conclusion, we have verified by optical techniques that symmetric-like silver nanoparticles embedded in a glass matrix suffer an anisotropic deformation upon Si-ion irradiation. The optical response has been simulated, finding that nanoparticles acquired an aligned prolate spheroidal shape, which explains the observed optical anisotropy along the direction of the Si-ion beam. As the irradiation fluence increases, the deformation rate also does, indicating a control of the deformation by varying the Si-ion fluence.

 We acknowledge the fruitful discussions with Ignacio L. Garz\'on.
  Partial financial support from CONACyT Grants No. 36651-E, 40122, and 42626-F, and DGAPA-UNAM Grants No.~IN101605 and IN-104303 is acknowledged. 
  K.  L\'opez, F.J. Jaimes, J.G. Morales, L. Rend\'on  and Dr. J. Arenas are also acknowledged. 

 \end{document}